# Near-Field Terahertz Nanoscopy of Coplanar Microwave Resonators

Xiao Guo,[1] Xin He,[2,3] Zach Degnan,[2] Bogdan C. Donose,[1] Karl Bertling,[1] Arkady Fedorov,[2,3] Aleksandar D. Rakić,[1,a)] and Peter Jacobson[2,b)]
[1)] *School of Information Technology and Electrical Engineering, The University of Queensland, Brisbane, QLD 4072, Australia*
[2)] *School of Mathematics and Physics, The University of Queensland, Brisbane, QLD 4072, Australia*
[3)] *ARC Centre of Excellence for Engineered Quantum Systems, Brisbane, QLD 4072, Australia*

Superconducting quantum circuits are one of the leading quantum computing platforms. To advance superconducting quantum computing to a point of practical importance, it is critical to identify and address material imperfections that lead to decoherence. Here, we use terahertz Scanning Near-field Optical Microscopy (SNOM) to probe the local dielectric properties and carrier concentrations of wet-etched aluminum resonators on silicon, one of the most characteristic components of the superconducting quantum processors. Using a recently developed vector calibration technique, we extract the THz permittivity from spectroscopy in proximity to the microwave feedline. Fitting the extracted permittivity to the Drude model, we find that silicon in the etched channel has a carrier concentration greater than buffer oxide etched silicon and we explore post-processing methods to reduce the carrier concentrations. Our results show that near-field THz investigations can be used to quantitatively evaluate and identify inhomogeneities in quantum devices.

The largest current impediment to the widespread implementation of solid-state quantum computation is decoherence.[1,2] For superconducting devices, decoherence is closely tied to how superconducting elements and microwave photons interact with the material environment.[3–5] To maintain quantum coherence, it is critical to optimize fabrication protocols and reduce material imperfections which generate losses.[6–8] Given the relatively simple design of superconducting coplanar waveguide resonators, they are an ideal experimental testbed for refining fabrication procedures and identifying loss channels.[9] Furthermore, since resonators and qubits share materials and fabrication procedures, imperfections identified as loss channels in resonators are probable qubit decoherence channels. This similarity means that processing strategies that reduce resonator losses (increase quality factor) can be rapidly translated to the fabrication of improved qubits.[10]

Our current understanding of losses in superconducting devices at millikelvin temperatures and low powers is based on the concept of tunneling in two-level systems (TLS). These systems, long studied in glasses and in Josephson junctions, are broadly thought to be related to light atoms transitioning between energetically equivalent positions at defect sites.[11–13] For thin-film devices, a major source of TLS is believed to originate from amorphous materials, variations in material stoichiometry, strain, and impurity segregation.[13,14] To reduce the influence of TLS on superconducting device performance multiple groups have turned to techniques such as deep reactive ion etching or wet-chemical treatments.[15–17] Recently, it was shown that thin $SiO_x$ layers formed after metal etching are a substantial contributor to TLS losses, and removing these layers with buffered oxide etch (BOE) greatly increases the quality factor.[18] However, our understanding of fabrication and post-processing steps contains substantial knowledge gaps. Frequently, fabrication protocols are varied and devices are tested at low temperatures to discern trends, or samples are subjected to destructive methods in preparation for measurements (e.g. ion milling for electron microscopy). These methods preclude the tracking of microscopic changes to the device after repeated procedures.

Scanning near-field optical microscopy (SNOM) is a rapidly evolving technique that enables far subwavelength imaging of the optical response.[22–24] Central to apertureless or scattering SNOM is light directed onto a metal-coated scanning probe tip, leading to the formation of a tightly localized electric field at the tip apex. This confined field enables the spatial resolution dependent only on the tip size and achieves nanoscale resolution for long wavelengths. In particular, THz SNOM[25–28] enables nanoscale imaging with the THz spectrum[29], which matches the fundamental energy and time scales in condensed matter systems.[30] Notable uses of THz SNOM include probing surface waves,[31] resolving femtosecond interlayer tunnelling processes,[32] mapping of carrier concentrations in integrated circuits,[21,33] and tracking phase transitions in correlated materials at the nanoscale.[34,35]

Here, we use THz SNOM to investigate the material properties and imperfections of wet-etched aluminum resonators on silicon in a non-invasive way after different fabrication steps. Combining THz time-domain spectroscopy with a scattering SNOM, we spatially resolve nanoscale THz features and collect broadband THz spectra. We find that the permittivity and carrier concentra-

[a)]
[b)] Author to whom correspondence should be addressed: [Aleksandar D. Rakić, a.rakic@uq.edu.au; Peter Jacobson, p.jacobson@uq.edu.au]

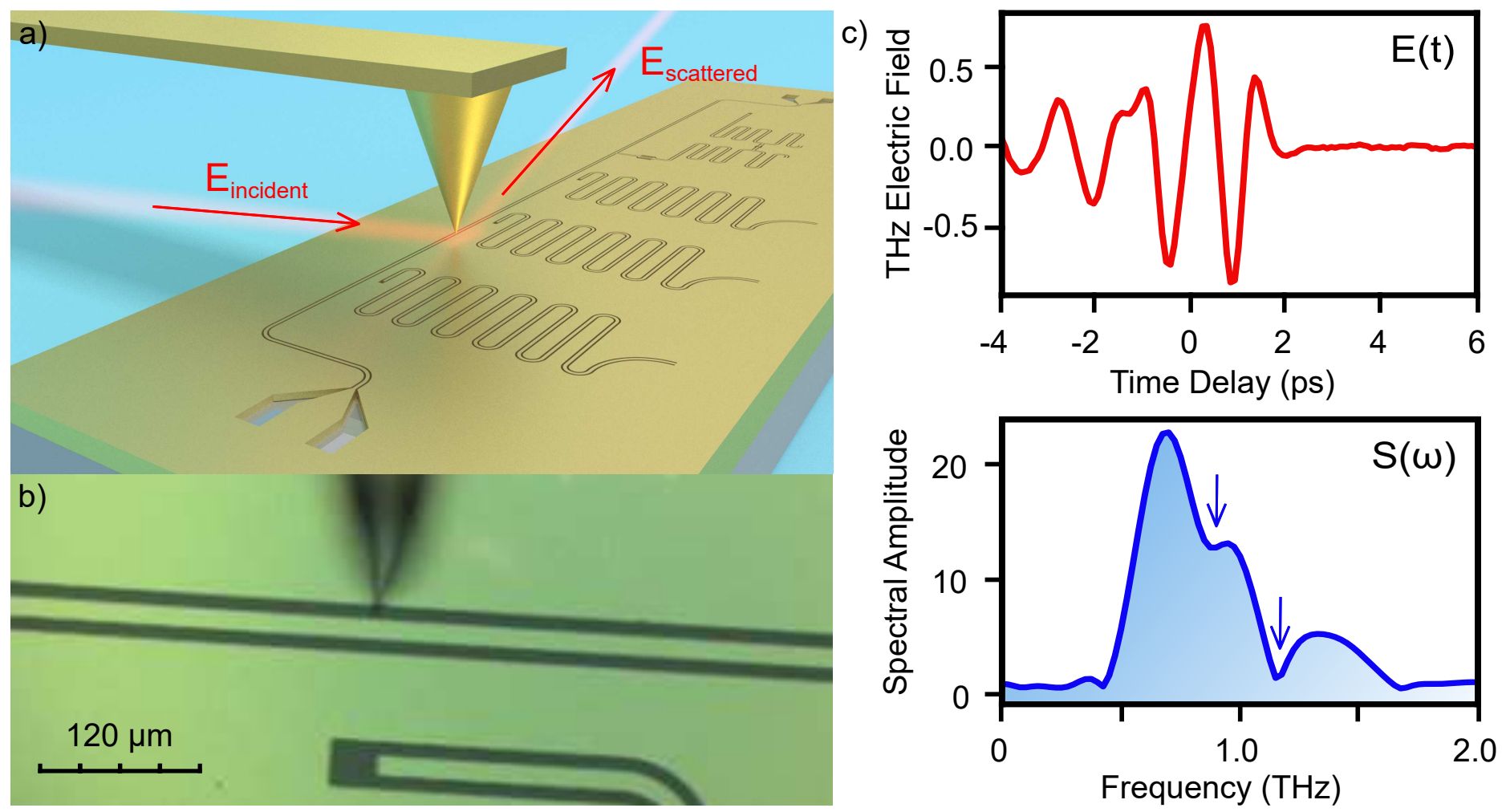

FIG. 1. (a) 3D schematic showing SNOM geometry and the resonator chip. (b) Optical micrograph of the region of interest, areas that are etched to expose the underlying silicon substrate appear dark. (c) Measured time-domain THz waveform, E(t), used for the experiments (top). The frequency-dependent THz spectrum, S($\omega$) is obtained by taking a fast Fourier transform of the time-domain waveform. Dips in S($\omega$) between 0.8 and 1.2 THz correspond to water vapor lines.[19–21]

tion of silicon in wet-etched regions are modified, leading to increased near-field THz scattering compared to BOE high-resistivity silicon. Our results show the power of combining microscopy with local THz spectroscopy and can be used to identify processing-induced inhomogeneities that limit device function.

High resistivity Si(100) (Topsil, floating zone grown, $\rho > 10$ k$\Omega \cdot$ cm) was prepared by etching in 5% BOE for 20 s to generate a hydrogen-passivated surface. The etched substrate was transferred to a Plassys MEB 550S electron beam evaporator and $\sim$ 85 nm of aluminum (1 nm/s by quartz microbalance) was deposited at room temperature. Coplanar waveguide resonators were fabricated in a process similar to Burnett et al.[16] The sample is spin-coated with AZ1512-HS resist and patterned using direct-write lithography (Heidelberg Instruments $\mu$PG 101). Once the pattern is defined and developed (AZ726), aluminum is selectively etched from exposed regions using an etchant containing 21% deionized (DI) water, 73% $H_3PO_4$, 3% acetic acid, and 3% $HNO_3$ by volume. When the etching is complete, the remaining photoresist is removed by submerging for 2 minutes in 60 °C VLSI acetone, followed by a 15 second rinse in VLSI isopropanol. The wafer is then dried with nitrogen gas.

All THz near-field measurements were performed under ambient conditions (40 – 60 % relative humidity with variation during measurements smaller than 4%) with a SNOM. It combines a broad-band THz-TDS system (TeraSmart, Menlo Co.) with a scattering-type SNOM (NeaSNOM, Neaspec GmbH), operated in tapping mode ($f_0 = 73.7$ kHz, A $\sim 110$ nm) with PtIr-coated AFM tips (25PtIr200B-H, Rocky Mountain Nanotechnology). For near-field measurements, sample topography, mechanical phase, and THz near-field data channels were collected.

Figure 1 (a) shows a 3D rendering of the SNOM experimental geometry and layout of the resonator chip. The SNOM tip was positioned above the 8 $\mu$m wide etched channel between the microwave transmission line and the ground plane (Figure 1 (b)), multiple locations were sampled at each post-processing step. Initial measurements were performed on a resonator chip where the circuit layout was defined by wet chemical etching before any post-processing procedures (oxide etching or undercuts). With the resonator chip loaded into the SNOM, broadband single-cycle pulses (0 − 6 THz) generated by a Fe-implanted InGaAs/InAlAs photoconductive antenna (TERA 15-TX-FC, MenloSystems) are focused onto the metallic AFM tip and the elastically forward-scattered field is detected by a low-temperature-grown InGaAs/InAlAs photoconductive antenna (TERA 15-RX-FC, Menlosystems). An optical scanning delay line is used to obtain the time-dependent THz scattering field E(t); a delay position with the maximum THz time-domain scattering amplitude was chosen for THz near-field imaging. THz scattering signals were demodulated with different harmonics ($n \leq 2$) of the probe oscillation frequency to obtain near-field signals $S_n(t)$. A Fourier transform was performed to obtain THz near-field spectra $S_n(\omega)$ (Figure 1 (c)).

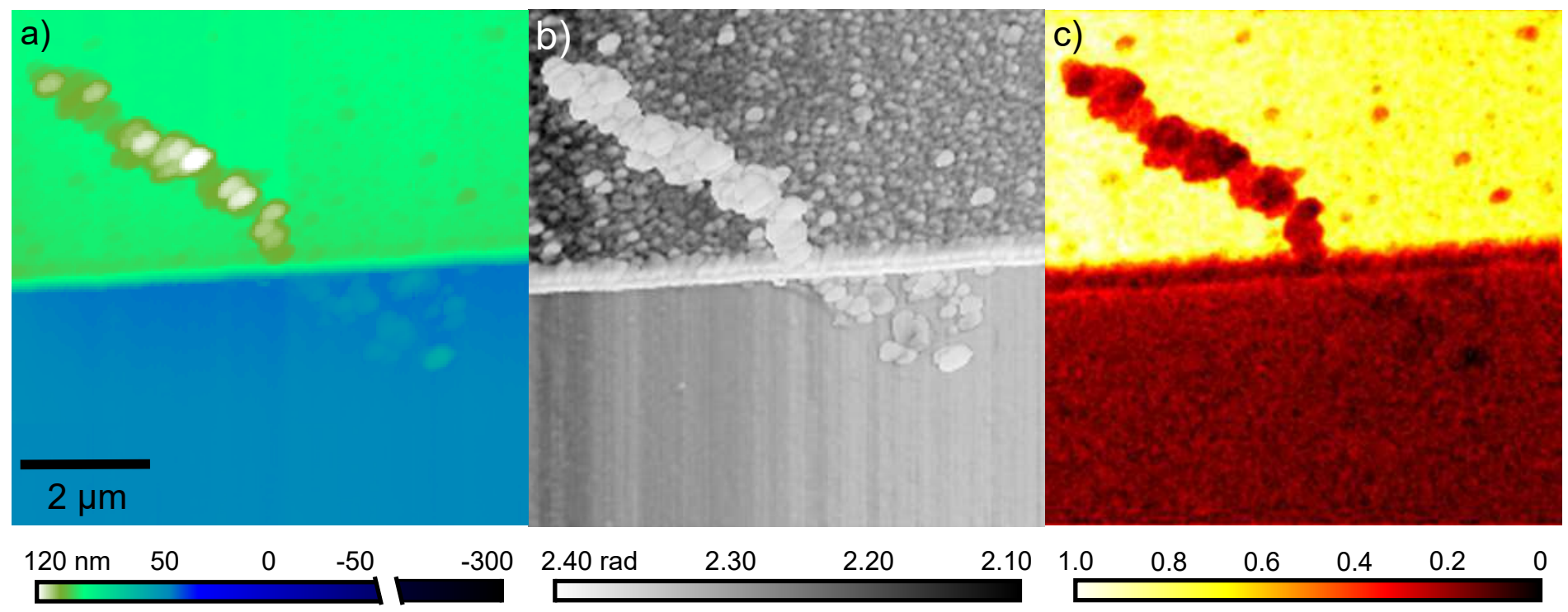

FIG. 2. (a, b) AFM topography, phase, and (c) THz SNOM imaging of the aluminum-silicon boundary region. The upper portion of the image corresponds to the aluminum film and the lower portion to the etched silicon channel. The feature crossing the boundary is residual photoresist from the lift-off process. The broadband THz near-field amplitude image, $S_2$, reveals the spatially varying electron-electron scattering efficiency.

Tapping mode AFM of the region straddling the aluminum and etched channel confirms that the aluminum film is 85 nm thick with a sharp transition between regions (Figure 2 (a)). This area contains an elongated structure immobilized on the surface that crosses from the aluminum into the etched channel. We attribute this feature to residual photoresist from the lift-off process and discuss this feature below. As the aluminum film and etched channel have a similar flat appearance in AFM topography, we turn to phase images to confirm the complete etching of the aluminum film in the channel (Figure 2 (b)). The etched region shows a uniform phase contrast, whereas the aluminum film has an inhomogeneous appearance consistent with a polycrystalline film capped by native oxides or hydroxides. This strong variation between regions in the phase image indicates that wet-etching has exposed the underlying silicon substrate.

Figure 2 (c) shows the second harmonic THz near-field response, simultaneously obtained with topography and phase, and shows a contrast between the aluminum film and the exposed silicon substrate related to the free carrier concentration of the different materials. While the THz response remains essentially unchanged across the etched channel; on the other side, the aluminum film (excluding the extended residue) contains regions with local reductions in the THz scattering amplitude. The extended residue has a spatially varying THz response indicating the composition of the residue is not uniform. Comparing with Figure 2 (b), diminished THz response on the aluminum correlates with bright features in phase imaging. As the THz response at localized sites bears similarity to parts of the extended residue, we attribute these THz dark spots to smaller regions of photoresist residue. Within the etched channel, enhanced THz scattering is observed in the silicon channel around 500 nm from the aluminum-silicon boundary, this feature is due to a surface plasmon polariton and will be commented on elsewhere.

To better correlate the scattered THz response with the local material properties, we turn to THz nanospectroscopy. Utilizing reference signals obtained from three standards: (1) gold mirror (Thorlabs, PF05–03-M03),(2) high-resistivity float zone silicon (Tydex, BS-HRFZ-SI-D50.8-T5), (3) p-doped silicon with known doping $2 \times 10^{16}$ cm$^{-3}$ (Bruker Nano Inc, SCM Sample), we calibrate the SNOM system response using a recently developed vector method.[33] This method allows us to extract the real and imaginary components of the complex dielectric permittivity at selected locations. Figure 3a shows the calibrated complex permittivity measured on silicon for four sequential sample preparations: high resistivity silicon after 20 seconds of 5% BOE prior to aluminum deposition (blue), the as-prepared resonator (red), after 22 seconds of 5% BOE (green), and after a 10 s $XeF_2$ exposure at 1.5 Torr (magenta). For patterned samples, the SNOM tip was positioned at the center of the etched channel 4 $\mu m$ from the aluminum film. The complex permittivity of the BOE-treated silicon reference (blue) varies little across the spectral range and the slight deviation from 0.85 to 1.2 THz in all spectra is due to absorption by humid air.[19–21] A strong THz water-vapour line around 1.15 THz is highlighted as the vertical broken line in Figure 3 (a). In contrast, the as-prepared resonators (red) and BOE etched resonators show a reduced real (increased imaginary) permittivity at the lower end of the spectral range. Finally, the $XeF_2$ exposed surface (magenta) shows a near-uniform permittivity across the spectral range and is most similar to the silicon before film deposition and fabrication.

To inspect the same microscopic region of the silicon-aluminum boundary after multiple post-processing treatments, a small aluminum notch defect was used to repro-ducibly position the tip over the region shown in Figure 2. AFM topography shows that after the BOE dip the exposed silicon remains flat (Figure 3 (b)), however, the aluminum surface becomes noticeably rougher (root-mean-square roughness: ± 11 nm). Furthermore, the transition between aluminum film and the etched channel has become more inhomogeneous, suggesting some polycrystalline grains have been removed during the etching. We do observe the removal of photoresist residue from the aluminum-silicon boundary, likely due to BOE attacking the underlying $SiO_x$ (or $AlO_x$) and dislodging the resist. We note that a shorter 2-second BOE dip did not appreciably affect the aluminum film or residues (not shown). While BOE is highly selective at removing $SiO_x$ compared to aluminum (or $AlO_x$), the increased roughness of the aluminum film is an undesired side effect. $XeF_2$ is a highly selective silicon etchant with a rapid etching rate. While not standard in resonator or qubit fabrication, $XeF_2$ was chosen because the photoresist mask had been removed at an earlier stage. After this processing step, the previously smooth silicon channel is significantly roughened with upwards of 300 nm of silicon removed in some areas while keeping the aluminum film remains intact. Due to the aggressive nature of $XeF_2$ etching, AFM imaging quality was reduced and imaging of the tip or multiple tip artifacts were observed as seen in Figure 3 (c).

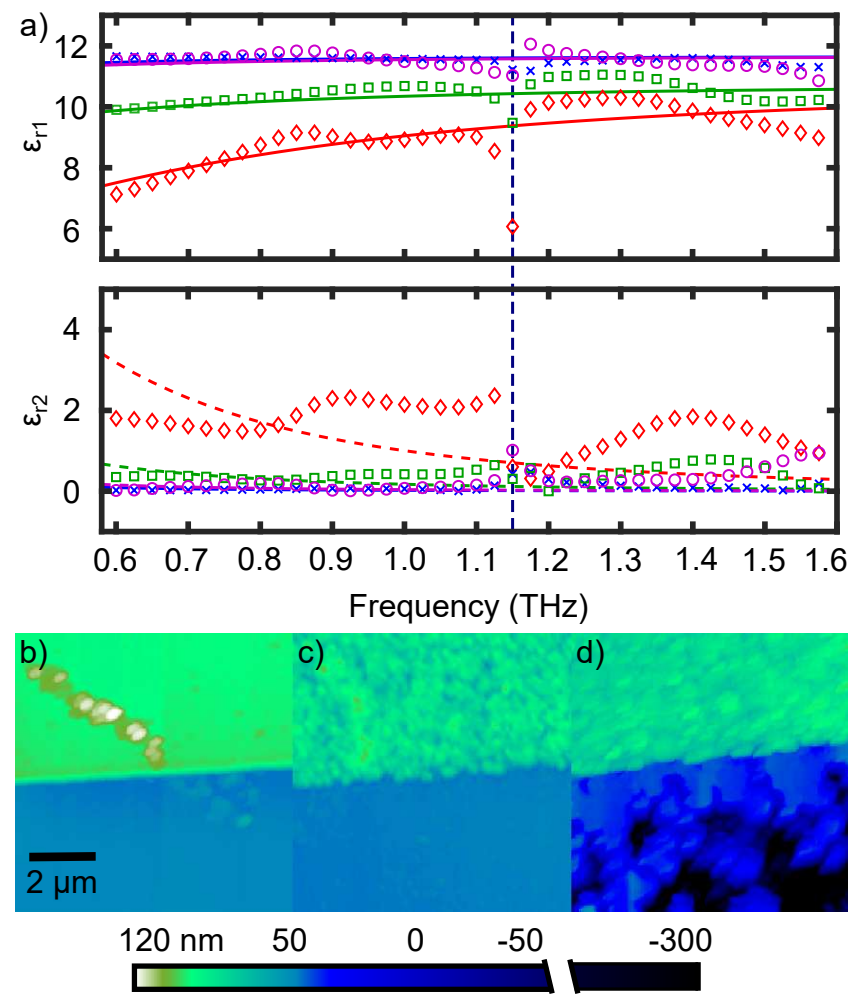

FIG. 3. (a) Extracted permittivity vs. Frequency for the silicon channel. $\epsilon_{r1}$ ( markers: measured, solid lines: fit) and $\epsilon_{r2}$ (markers: measured, dashed lines: fit), for Step 0: BOE Si reference (blue, ×), Step 1: as-prepared resonators (red, ⋄), Step 2: 22-sec BOE Etch (green, □), Step 3: $XeF_2$ Etch (magenta, ∘) (vertical broken line, THz water-vapour line around 1.15 THz). (b,c,d) AFM topography for Step 1: as-prepared, Step 2: BOE-treated and Step 3: $XeF_2$-treated resonators.

Our THz permittivity measurements in Figure 3 (a) provide insight into the local electronic properties since the frequency-dependent conductivity is directly proportional to the imaginary part of permittivity, $\epsilon_{r2}$. To extract the carrier concentration, we simultaneously fit the real and imaginary components of the permittivity at each processing step using the Drude model (solid, dashed lines Figure 3 (a)). Table I contains the extracted carrier concentrations ($N_c$) from the Drude model fits[36] with one standard error. After patterning and etching the aluminum film that defines the resonators, the carrier concentration within the silicon channel is substantially increased compared to BOE etched silicon. Exposing the as-prepared resonator chip to BOE decreases the carrier concentration by an order of magnitude, we attribute this reduction to the removal of charged impurities in the thin $SiO_x$ layer.[37,38] A similar order of magnitude drop in carrier concentration is observed after $XeF_2$ exposure, returning the silicon to a nominally undoped state comparable to the high-resistivity silicon substrate before metal deposition.

Our results show that high-resistivity silicon substrates are unintentionally doped during the resonator fabrication process and device post-processing can reduce the doping within the exposed channel. While BOE treatment removes excess carriers (and photoresist residue) from the silicon channel, the carrier concentration is still greater than the high-resistivity silicon starting point. This indicates that only a fraction of the excess carriers can be attributed to defect structures within the thin $SiO_x$ layer.[37,38] Exposing our device to $XeF_2$ removes material and charge carriers, indicating that unintentional dopants are localized to the near-surface region. From the chemicals used to fabricate our devices, phosphoric acid ($H_3PO_4$) is the major component of the aluminum etchant and has previously been employed to dope silicon.[39] The dopant distribution, gauged by the THz scattering amplitude, appears uniform on micrometer length scales for as-prepared and BOE etched samples. We note that the same THz SNOM was previously used by some of the authors to visualize n- and p-doped regions in SRAM devices and extract carrier concentrations on the order of $10^{16}$ $cm^{-3}$ and $10^{17}$ $cm^{-3}$, in agreement with the manufacturer's specifications and literature.[21,33] The highest carrier concentration observed in our resonator sample ($10^{16}$ $cm^{-3}$) overlaps with this lower bound, providing confidence in the ability of THz SNOM to detect spatially varying dopant distributions.

Surface treatments after metal etching are now standard in fabrication protocols for superconducting devices.[16,40] Reports on these treatments are often focused on the presence and abatement of surface oxides, undoubtedly a major contributor to resonator losses in the low-temperature and low-power regime.[18] The work here presents evidence for a previously unconsidered effect — fabrication-induced doping of silicon. Charge carriers in proximity to the microwave feedlines will move in re-sponse to an applied microwave signal and are a potential source of microwave-induced quasiparticles, particularly in regions of high electric fields.[41] Furthermore, doping of the near-surface region alters the dielectric properties which may affect the coupling strength between the microwave feedline and the resonators. Scattering-SNOM can operate at liquid helium temperatures, which opens the door to dielectric characterization nearer to the device operating temperature (mK).[42,43] For a dopant profile localized in the surface and near-surface region, methods such as X-ray photoelectron spectroscopy or secondary ion mass spectrometry (SIMS) could be used to determine the chemical identity and spatial distribution of the dopants.[44,45] In particular, SIMS depth profiling could identify the optimal etching depth needed to remove surface doping while minimizing structural damage to the substrate.

TABLE I. Extracted doping variation $N_c$ for different surface treatments from measured plasma frequency $\omega_p$ assuming effective carrier mass $m^*$. (p-type: $m^* = 0.37m_0$, n-type: $m^* = 0.26m_0$, $m_0$ is free-electron mass)

| Step | Treatment | $\omega_p$ (THz) | $N_c$ (cm$^{-3}$), $m^* = 0.37m_0$ | $N_c$ (cm$^{-3}$), $m^* = 0.26m_0$ |
|---|---|---|---|---|
| 0 | BOE Si Reference | $0.378 \pm 0.031$ | $(6.541 \pm 1.015) \times 10^{14}$ | $(4.460 \pm 0.576) \times 10^{14}$ |
| 1 | As-prepared Resonators | $1.614 \pm 0.051$ | $(1.195 \pm 0.093) \times 10^{16}$ | $(8.398 \pm 0.655) \times 10^{15}$ |
| 2 | 22-sec BOE Etch | $0.879 \pm 0.076$ | $(3.458 \pm 0.588) \times 10^{15}$ | $(2.494 \pm 0.413) \times 10^{15}$ |
| 3 | $XeF_2$ Etch | $0.453 \pm 0.069$ | $(9.408 \pm 2.657) \times 10^{14}$ | $(6.616 \pm 1.869) \times 10^{14}$ |

We have used THz SNOM and nanospectroscopy to quantify the effect of post-processing protocols on coplanar waveguide resonators. From spectrally resolved permittivity measurements, we find that etched regions of the silicon substrate contain excess carriers and that these carriers can be removed to varying degrees by chemical etching. Our work shows that THz SNOM can be used to rapidly screen materials and inform material processing decisions critical to the fabrication of quantum circuitry.

## ACKNOWLEDGMENTS

The authors acknowledge the Traditional Owners and their custodianship of the lands on which UQ operates. We pay our respects to their Ancestors and their descendants, who continue cultural and spiritual connections to Country. Financial support was provided by the Australian Research Council (DP200101948 and DP210103342) and the ARC Centre of Excellence for Engineered Quantum Systems (EQUS, CE170100009). The authors acknowledge the facilities, and the scientific and technical assistance, of the Microscopy Australia Facility at the Centre for Microscopy and Microanalysis, The University of Queensland. Part of this work used the Queensland node of the NCRIS-enabled Australian National Fabrication Facility (ANFF).

## DATA AVAILABILITY STATEMENT

The data that support the findings of this study are available from the corresponding author upon reasonable request.

[1] J. M. Martinis, "Qubit metrology for building a fault-tolerant quantum computer," npj Quantum Inf. **1**, 1–3 (2015).

[2] N. P. de Leon, K. M. Itoh, D. Kim, K. K. Mehta, T. E. Northup, H. Paik, B. S. Palmer, N. Samarth, S. Sangtawesin, and D. W. Steuerman, "Materials challenges and opportunities for quantum computing hardware," Science **372** (2021).

[3] J. M. Martinis, K. B. Cooper, R. McDermott, M. Steffen, M. Ansmann, K. D. Osborn, K. Cicak, S. Oh, D. P. Pappas, R. W. Simmonds, and C. C. Yu, "Decoherence in josephson qubits from dielectric loss," Phys. Rev. Lett. **95**, 210503 (2005).

[4] W. D. Oliver and P. B. Welander, "Materials in superconducting quantum bits," MRS Bull. **38**, 816–825 (2013).

[5] C. E. Murray, "Material matters in superconducting qubits," arXiv preprint arXiv:2106.05919 (2021).

[6] J. Gao, M. Daal, A. Vayonakis, S. Kumar, J. Zmuidzinas, B. Sadoulet, B. A. Mazin, P. K. Day, and H. G. Leduc, "Experimental evidence for a surface distribution of two-level systems in superconducting lithographed microwave resonators," Appl. Phys. Lett. **92**, 152505 (2008).

[7] A. Megrant, C. Neill, R. Barends, B. Chiaro, Y. Chen, L. Feigl, J. Kelly, E. Lucero, M. Mariantoni, P. J. J. O'Malley, D. Sank, A. Vainsencher, J. Wenner, T. C. White, Y. Yin, J. Zhao, C. J. Palmstrøm, J. M. Martinis, and A. N. Cleland, "Planar superconducting resonators with internal quality factors above one million," Appl. Phys. Lett. **100**, 113510 (2012).

[8] C. J. K. Richardson, N. P. Siwak, J. Hackley, Z. K. Keane, J. E. Robinson, B. Arey, I. Arslan, and B. S. Palmer, "Fabrication artifacts and parallel loss channels in metamorphic epitaxial aluminum superconducting resonators," Supercond. Sci. Technol. **29**, 064003 (2016).

[9] C. T. Earnest, J. H. Béjanin, T. G. McConkey, E. A. Peters, A. Korinek, H. Yuan, and M. Mariantoni, "Substrate surface engineering for high-quality silicon/aluminum superconducting resonators," Supercond. Sci. Technol. **31**, 125013 (2018).

[10] J. J. Burnett, A. Bengtsson, M. Scigliuzzo, D. Niepce, M. Kudra, P. Delsing, and J. Bylander, "Decoherence benchmarking of superconducting qubits," npj Quantum Inf. **5**, 1–8 (2019).

[11] P. W. Anderson, B. I. Halperin, and C. M. Varma, "Anomalous low-temperature thermal properties of glasses and spin glasses," Philos. Mag. **25**, 1–9 (1972).

[12] G. J. Grabovskij, T. Peichl, J. Lisenfeld, G. Weiss, and A. V. Ustinov, "Strain tuning of individual atomic tunneling systems detected by a superconducting qubit," Science **338**, 232–234 (2012).

[13] C. Müller, J. H. Cole, and J. Lisenfeld, "Towards understanding two-level-systems in amorphous solids: insights from quantum circuits," Rep. Progr. Phys. **82**, 124501 (2019).

[14] A. P. Paz, I. V. Lebedeva, I. V. Tokatly, and A. Rubio, "Identification of structural motifs as tunneling two-level systems in amorphous alumina at low temperatures," Phys. Rev. B **90**, 224202 (2014).

[15]A. Bruno, G. De Lange, S. Asaad, K. L. Van Der Enden, N. K. Langford, and L. DiCarlo, "Reducing intrinsic loss in superconducting resonators by surface treatment and deep etching of silicon substrates," Appl. Phys. Lett. **106**, 182601 (2015).

[16]J. Burnett, A. Bengtsson, D. Niepce, and J. Bylander, "Noise and loss of superconducting aluminium resonators at single photon energies," in *J. Phys. Conf. Ser.*, Vol. 969 (IOP Publishing, 2018) p. 012131.

[17]G. Calusine, A. Melville, W. Woods, R. Das, C. Stull, V. Bolkhovsky, D. Braje, D. Hover, D. K. Kim, X. Miloshi, D. Rosenberg, A. Sevi, J. L. Yoder, E. Dauler, and W. D. Oliver, "Analysis and mitigation of interface losses in trenched superconducting coplanar waveguide resonators," Appl. Phys. Lett. **112**, 062601 (2018).

[18]M. V. P. Altoé, A. Banerjee, C. Berk, A. Hajr, A. Schwartzberg, C. Song, M. A. Ghadeer, S. Aloni, M. J. Elowson, J. M. Kreikebaum, E. K. Wong, S. Griffin, S. Rao, A. Weber-Bargioni, A. M. Minor, D. I. Santiago, S. Cabrini, I. Siddiqi, and D. F. Ogletree, "Localization and reduction of superconducting quantum coherent circuit losses," arXiv preprint arXiv:2012.07604 (2020).

[19]M. Van Exter, C. Fattinger, and D. Grischkowsky, "Terahertz time-domain spectroscopy of water vapor," Opt. Lett. **14**, 1128–1130 (1989).

[20]Y. Yang, M. Mandehgar, and D. Grischkowsky, "Determination of the water vapor continuum absorption by THz-TDS and molecular response theory," Opt. Express **22**, 4388–4403 (2014).

[21]N. A. Aghamiri, F. Huth, A. J. Huber, A. Fali, R. Hillenbrand, and Y. Abate, "Hyperspectral time-domain terahertz nano-imaging," Opt. Express **27**, 24231–24242 (2019).

[22]F. Keilmann, A. J. Huber, and R. Hillenbrand, "Nanoscale conductivity contrast by scattering-type near-field optical microscopy in the visible, infrared and THz domains," J. Infrared Millim. Terahertz Waves **30**, 1255–1268 (2009).

[23]X. Chen, D. Hu, R. Mescall, G. You, D. Basov, Q. Dai, and M. Liu, "Modern scattering-type scanning near-field optical microscopy for advanced material research," Adv. Mater. , 1804774 (2019).

[24]Z. Yao, S. Xu, D. Hu, X. Chen, Q. Dai, and M. Liu, "Nanoimaging and nanospectroscopy of polaritons with time resolved s-SNOM," Adv. Opt. Mater. , 1901042 (2019).

[25]H.-G. von Ribbeck, M. Brehm, D. van der Weide, S. Winnerl, O. Drachenko, M. Helm, and F. Keilmann, "Spectroscopic THz near-field microscope," Opt. Express **16**, 3430–3438 (2008).

[26]A. Rakić, T. Taimre, K. Bertling, Y. Lim, P. Dean, A. Valavanis, and D. Indjin, "Sensing and imaging using laser feedback interferometry with quantum cascade lasers," Appl. Phys. Rev. **6**, 021320 (2019).

[27]E. A. Pogna, M. Asgari, V. Zannier, L. Sorba, L. Viti, and M. S. Vitiello, "Unveiling the detection dynamics of semiconductor nanowire photodetectors by terahertz near-field nanoscopy," Light Sci. Appl. **9**, 1–12 (2020).

[28]P. Rubino, J. Keeley, N. Sulollari, A. D. Burnett, A. Valavanis, I. Kundu, M. C. Rosamond, L. Li, E. H. Linfield, A. G. Davies, J. E. Cunningham, and P. Dean, "All-electronic phase-resolved THz microscopy using the self-mixing effect in a semiconductor laser," ACS Photonics **8**, 1001–1006 (2021).

[29]R. Lewis, "A review of terahertz detectors," J. Phys. D: Appl. Phys. **52**, 433001 (2019).

[30]R. A. Lewis, "Physical phenomena in electronic materials in the terahertz region," Proc. IEEE **95**, 1641–1645 (2007).

[31]T. V. A. G. de Oliveira, T. Nörenberg, G. Álvarez-Pérez, L. Wehmeier, J. Taboada-Gutiérrez, M. Obst, F. Hempel, E. J. Lee, J. M. Klopf, I. Errea, A. Y. Nikitin, S. C. Kehr, P. Alonso-Gonzalez, and L. M. Eng, "Nanoscale-confined terahertz polaritons in a van der Waals crystal," Adv. Mater. **33**, 2005777 (2021).

[32]M. Plankl, P. F. Junior, F. Mooshammer, T. Siday, M. Zizlsperger, F. Sandner, F. Schiegl, S. Maier, M. A. Huber, M. Gmitra, J. Fabian, J. L. Boland, T. L. Cocker, and R. Huber, "Subcycle contact-free nanoscopy of ultrafast interlayer transport in atomically thin heterostructures," Nat. Photonics , 1–7 (2021).

[33]X. Guo, K. Bertling, and A. D. Rakić, "Optical constants from scattering-type scanning near-field optical microscope," Appl. Phys. Lett. **118**, 041103 (2021).

[34]H. T. Stinson, A. Sternbach, O. Najera, R. Jing, A. S. Mcleod, T. V. Slusar, A. Mueller, L. Anderegg, H. T. Kim, M. Rozenberg, and D. N. Basov, "Imaging the nanoscale phase separation in vanadium dioxide thin films at terahertz frequencies," Nat. Commun. **9**, 1–9 (2018).

[35]C. Chen, S. Chen, R. P. Lobo, C. Maciel-Escudero, M. Lewin, T. Taubner, W. Xiong, M. Xu, X. Zhang, X. Miao, P. Li, and R. Hillenbrand, "Terahertz nanoimaging and nanospectroscopy of chalcogenide phase-change materials," ACS Photonics **7**, 3499–3506 (2020).

[36]M. Van Exter and D. Grischkowsky, "Carrier dynamics of electrons and holes in moderately doped silicon," Phys. Rev. B **41**, 12140 (1990).

[37]H. Angermann, T. Dittrich, and H. Flietner, "Investigation of native-oxide growth on HF-treated Si (111) surfaces by measuring the surface-state distribution," Appl. Phys. A **59**, 193–197 (1994).

[38]V. Schmidt, S. Senz, and U. Gösele, "Influence of the Si/$SiO_2$ interface on the charge carrier density of Si nanowires," Appl. Phys. A **86**, 187–191 (2007).

[39]D. S. Kim, M. M. Hilali, A. Rohatgi, K. Nakano, A. Hariharan, and K. Matthei, "Development of a phosphorus spray diffusion system for low-cost silicon solar cells," J. Electrochem. Soc. **153**, A1391 (2006).

[40]D. Kalacheva, G. Fedorov, A. Kulakova, J. Zotova, E. Korostylev, I. Khrapach, A. V. Ustinov, and O. V. Astafiev, "Improving the quality factor of superconducting resonators by post-process surface treatment," in *AIP Conf. Proc.*, Vol. 2241 (AIP Publishing LLC, 2020) p. 020018.

[41]P. J. De Visser, J. J. A. Baselmans, S. J. C. Yates, P. Diener, A. Endo, and T. M. Klapwijk, "Microwave-induced excess quasiparticles in superconducting resonators measured through correlated conductivity fluctuations," Appl. Phys. Lett. **100**, 162601 (2012).

[42]H. U. Yang, E. Hebestreit, E. E. Josberger, and M. B. Raschke, "A cryogenic scattering-type scanning near-field optical microscope," Rev. Sci. Instrum. **84**, 023701 (2013).

[43]D. Lang, J. Döring, T. Nörenberg, Á. Butykai, I. Kézsmárki, H. Schneider, S. Winnerl, M. Helm, S. C. Kehr, and L. M. Eng, "Infrared nanoscopy down to liquid helium temperatures," Rev. Sci. Instrum. **89**, 033702 (2018).

[44]P. C. Zalm, "Ultra shallow doping profiling with SIMS," Rep. Progr. Phys. **58**, 1321 (1995).

[45]S. Eswara, A. Pshenova, E. Lentzen, G. Nogay, M. Lehmann, A. Ingenito, Q. Jeangros, F.-J. Haug, N. Valle, P. Philipp, A. Hessler-Wyser, and T. Wirtz, "A method for quantitative nanoscale imaging of dopant distributions using secondary ion mass spectrometry: an application example in silicon photovoltaics," MRS Commun. **9**, 916–923 (2019).

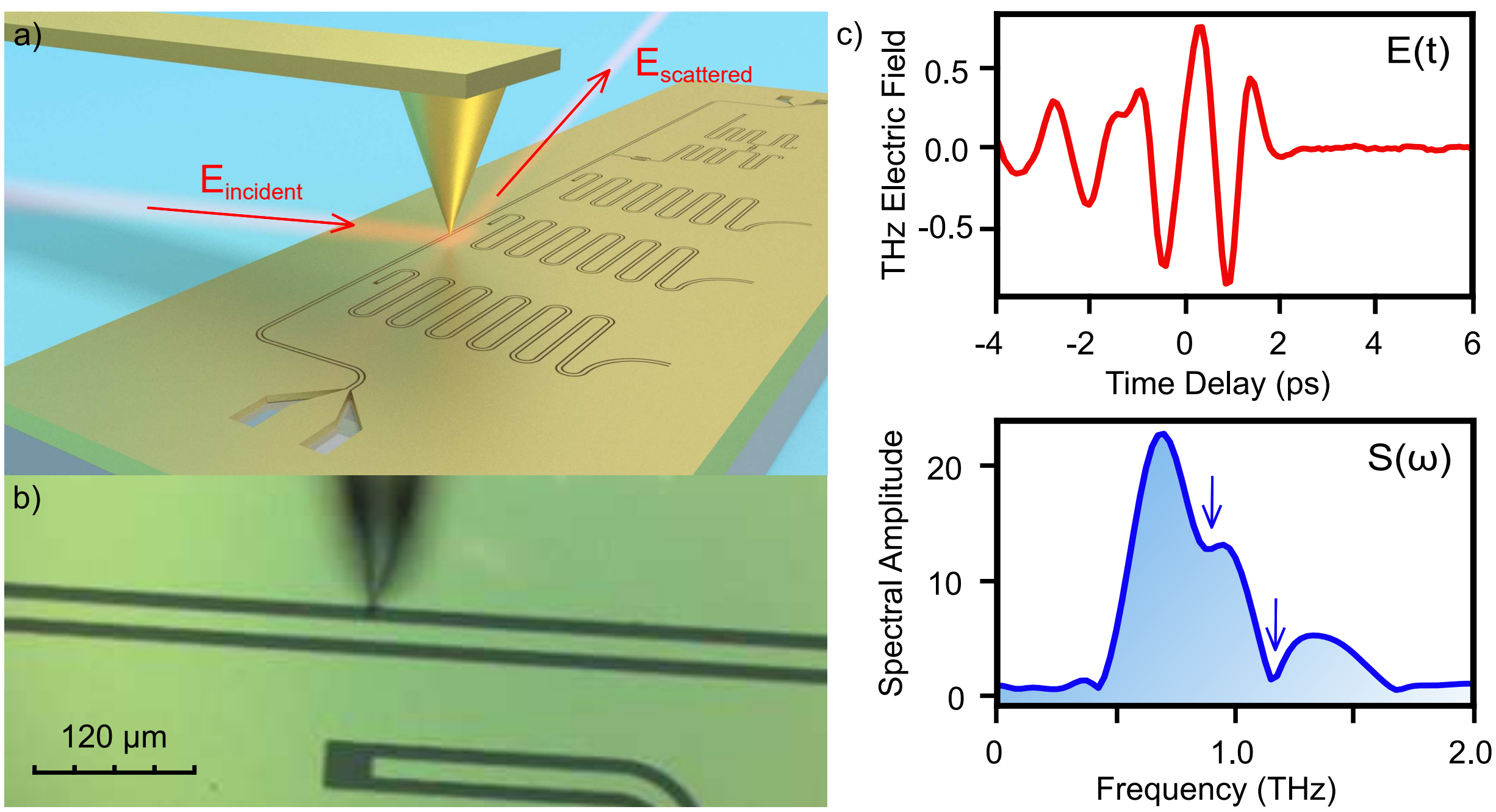
a)
$E_{scattered}$
$E_{incident}$
b)
120 µm
c)
E(t)
THz Electric Field
0.5
0.0
-0.5
-4
-2
0
2
4
6
Time Delay (ps)
S(ω)
Spectral Amplitude
20
10
0
0
1.0
2.0
Frequency (THz)

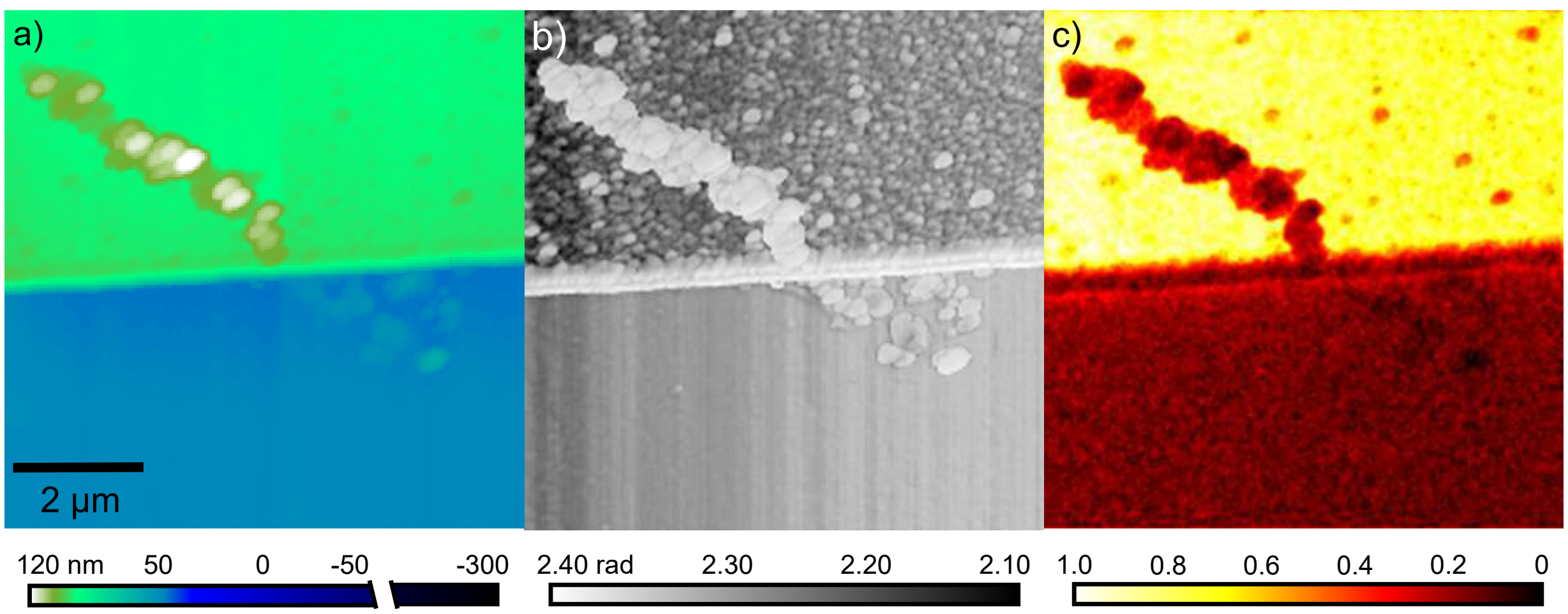
a)
b)
c)
2 µm
120 nm 50 0 -50 -300
2.40 rad 2.30 2.20 2.10
1.0 0.8 0.6 0.4 0.2 0

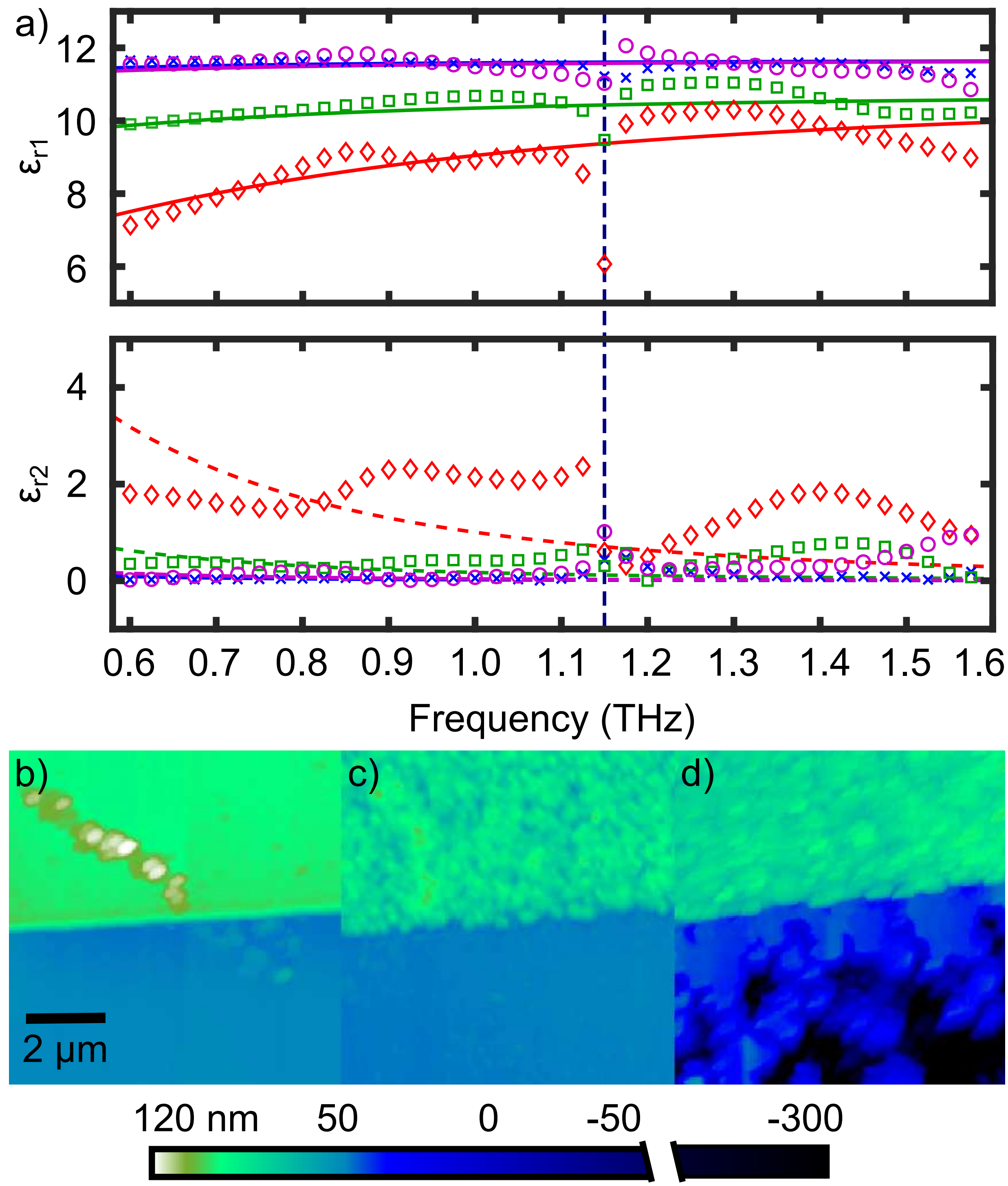
a)
12
10
8
6
$\varepsilon_{r1}$
4
2
0
$\varepsilon_{r2}$
0.6
0.7
0.8
0.9
1.0
1.1
1.2
1.3
1.4
1.5
1.6
Frequency (THz)
b)
c)
d)
2 μm
120 nm
50
0
-50
-300